\documentstyle[prl,aps]{revtex}
\begin{document}
\draft
\title{Measurement of the Probability Distribution of Total 
Transmission in Random Waveguides}
\author{M. Stoytchev and A. Z. Genack}
\address{New York State Center for Advanced Technology 
for Ultrafast Photonic Materials and Applications, Dept. 
of Physics, Queens College of CUNY, Flushing, NY 11367}

\maketitle

\begin{abstract}
Measurements have been made of the probability distribution of total 
transmission of microwave radiation in waveguides filled with randomly 
positioned scatterers which would have values of the dimensionless 
conductance $g$ near unity. The distributions are markedly non-Gaussian 
and have exponential tails. The measured distributions are accurately 
described by diagrammatic and random matrix calculations carried out for 
nonabsorbing samples in the limit $g \gg 1$  when $g$ is expressed in 
terms of the variance of the distribution, which equals the degree of 
long-range intensity correlation across the output face of the sample.
\end{abstract}
\pacs{42.25.Bs, 42.68.Mj, 41.20.Jb}
%\narrowtext
%\pagebreak

Nonlocal correlation in the flux transmitted through mesoscopic samples 
leads to enhanced fluctuations of local and spatially averaged 
transmission for both classical and quantum  waves.\cite{Sheng90,Alt91} 
Such fluctuations increase dramatically as the ensemble average of the 
dimensionless conductance, $g$, approaches unity. Low values of $g$ can 
be achieved in quasi-one-dimensional samples such as conducting wires 
or multimode waveguides with lengths much greater than the transverse 
dimensions. In this Letter, we report measurements of the probability 
distribution of total transmission of microwave radiation in long 
waveguides filled with randomly positioned scatterers which in the 
absence of absorption would have values of $g$ near unity. The 
distributions observed are markedly non-Gaussian. They are compared to 
recent diagrammatic and random-matrix calculations for nonabsorbing 
samples in the limit $g \gg 1$ \cite{vanRos95,Kog95}. This is done by 
reexpressing the distribution, which is a function of the single 
parameter g, as a function of the variance of the normalized 
transmission using the relation between these parameters. This result 
is in excellent agreement with the measured transmission distributions 
and indicates that the variance of the normalized transmission, which 
equals the degree of long-range intensity correlation across the output 
face of the sample, is the essential parameter describing fluctuations 
in random media.

Key transmission quantities in order of increasing spatial averaging are 
the intensity, $T_{ab}$, which is the transmission coefficient for 
incoming mode $a$ into mode $b$, the total transmission for incoming mode 
$a$, $T_a = \sum_{b}T_{ab} \sim \ell/L$, and the total transmittance 
$T = \sum_{ab}T_{ab} \sim N\ell/L$, where $\ell$ is the transport mean 
free path, $L$ is the sample length, and $N$ is the number of modes. The 
total transmittance is equivalent to the dimensionless conductance in 
electronic systems, $T = G/(e^2\!/h)$, where $G$ is the conductance 
\cite{Land70} and $g\, =\, <T>\, =\, N\ell /L$. Though the variances of 
the transmission quantities normalized to their ensemble average values, 
$s_{ab} = T_{ab}/\!\!\!<\!\!T_{ab}\!\!>$, $s_a = T_a/\!\!\!<\!\!T_a\!\!>$ 
and $s = T/\!\!\!<\!\!T\!\!>$, are reduced as the extent of spatial 
averaging increases, fluctuations in these quantities do not self average, 
as they would if spatial correlation were absent. To leading order in $1/g$, 
the enhancement of the variances of $s_{ab}$, $s_a$, and $s$ arising from 
nonlocal correlation is 1, $L/\ell$ \cite{Steph87}, and $(L/\ell)^2$ 
\cite{Lee85,Alt85}, respectively,  which results in values of the variances 
of 1, $1/g$, and $(1/g)^2$. \cite{Feng88,Mello88}

To examine the scaling and the universality of transport, it is important 
to measure the full distribution of key transmission quantities as the 
sample size, and hence $g$, changes. In previous work, nonlocal correlation 
has been shown to lead to higher probabilities at large values of the 
intensity, leading to a deviation from negative exponential statistics for 
polarized microwave radiation when $g \sim 10$ \cite{Gar89,Gen93}, as well 
as to discernable deviations from a Gaussian distribution and enhanced 
variance for the total optical transmission when $g > 10^3$ \cite{deBoer94}.

Recently, an expression for $P(s_a)$ in terms of $g$ for nonabsorbing samples 
was obtained by Nieuwenhuizen and van Rossum using diagrammatic techniques 
combined with random matrix theory \cite{vanRos95} and subsequently by Kogan 
and Kaveh within the framework of random matrix theory \cite{Kog95}. The 
diagrammatic calculations neglect some terms of order higher than $1/g$, 
whereas computations based on random matrix theory neglect sample-to-sample 
fluctuations in the probability distributions of eigenvalues of the 
transmission matrix and are expected to be accurate only to order $1/g$. 
More recently, van Langen, Brouwer and Beenakker carried out a 
nonperturbative calculation of the total transmission distribution in the 
absence of absorption \cite{vanLang96}. An analytic solution is obtained for 
the case in which time reversal invariance is broken $(\beta   = 2)$ but not 
for the case  of time reversal symmetry $(\beta = 1)$ considered here. 
However, good agreement is found between the $\beta$-independent result for 
$P(s_a)$ obtained in Refs.\  \cite{vanRos95,Kog95} and the result for 
$\beta = 2$ in Ref.\ \cite{vanLang96} for $g \geq 10$.

The distribution of total transmission has been measured previously by de 
Boer et al. in optical measurements in slabs of titania particles 
\cite{deBoer94}. Samples with $g > 10^3$ were studied and the distribution 
was found to be Gaussian to within $1 \%$. A measure of the deviations of 
the distribution from a Gaussian is the value of the third cumulant 
$<\!\!s_a^3\!\!>_c$ which gives the skewness of the distribution and 
vanishes for a Gaussian distribution. For the samples studied, 
$<\!\!s_a^3\!\!>_c$ was of order of $10^{-6}$. It was found that 
$<\!\!s_a^3\!\!>_c = \gamma _g <\!\!s_a^2\!\!>_c^2$ with $\gamma _g = 2.9 
\pm 0.4$ which is consistent with the value calculated for Gaussian beam 
excitation of 3.20 \cite{vanRos95}.

In the present work, low values of $g$ are achieved by placing the sample 
in a cylindrical copper tube in order to restrict transverse diffusion and 
thus the number of modes $N$. The samples consist of randomly positioned 
polystyrene spheres with diameters of 1.27 cm at a volume filling fraction 
$f = 0.55$. Transmission spectra were taken at tube diameters of 7.5 and 
5.0 cm and various sample lengths in the frequency range 16.8 - 17.8 GHz. 
The microwave radiation is coupled to the sample by a 0.4 cm wire antenna 
placed 0.5 cm from the front surface of the sample. The frequency is 
incremented in 4 MHz steps. The sample tube is rotated between successive 
measurements to produce new scatterer configurations. The total 
transmission is measured by use of a single Schottky diode detector 
positioned inside an integrating sphere which rotates at 2 Hz around the 
sample axis. The integrating sphere has a diameter of 40 cm and is 
comprised of two concentric plastic spherical shells separated by a 
distance of 2 cm. The outer shell is covered with aluminum foilt to form 
an irregular reflecting surface. The region between the shells is filled 
with thin-walled aluminum cylinders with diameters of 0.75 cm and typical 
lengths of 1 cm. The cylinders tumble as the integrating sphere rotates, 
resulting in fluctuations of the intensity at the detector with a 
correlation time of $\sim$ 2 ms for the sample with a length of 100 cm. 
The signal is averaged for 1 s at each frequency, giving an uncertainty of 
$2.5\ \%$ in the measurement of transmission. The signal is normalized by 
its ensemble average to give $s_a$. The transmission distributions $P(s_a)$ 
are obtained by using the data from at least 1000 sample configurations. 
Distributions obtained using different intervals of the frequency range 
coincide within experimental error. In the frequency range of the 
measurements, $\ell \approx 5$ cm and $N = k^2d^2/8 \approx 200$ and 90 for 
samples in tubes with diameters $d\, =\, 7.5$ and 5.0 cm, respectively. The 
wave number $k = 2\pi/\lambda = 2\pi\nu n/c$ is calculated using an 
effective medium index of refraction $n \approx 1.4$.

The transmission distributions for three samples with dimensions (a) 
$d = 7.5$ cm, $L = 66.7$ cm, (b) $d = 5.0$ cm, $L = 50.0$ cm, and (c) 
$d = 5.0$ cm, $L = 200$ cm are shown in Fig. 1. In the absence of 
absorption, the dimensionless conductance for these samples without 
localization corrections, $g = N\ell /L$, would be approximately 15.0, 
9.0, and 2.25 for samples $a$, $b$, and $c$, respectively. The 
distribution broadens and the deviation from a Gaussian becomes more 
pronounced as either the sample length increases or the cross-sectional 
area decreases. A value of $<\!\!s_a^3\!\!>_c$ as large as $0.112\, \pm\, 
0.003$ is observed for sample c. Deviations from a Gaussian distribution 
in the tail of the distribution for this sample can be seen in the 
semilog plot of $P(s_a)$ in Fig.\ 2. For values of $s_a\, \geq\, 2$ the 
distribution is nearly exponential. 

In Fig. 3, we present a plot of $<\!\!s_a^3\!\!>_c$ versus 
$<\!\!s_a^2\!\!>_c^2$. The solid line is a least square linear fit to the 
data which gives $\gamma = 2.38\, \pm\, 0.05$. Within experimental error 
this equals the value $\gamma = 2.40$ calculated for an incident plane 
wave in the lowest order of a diagrammatic perturbation expansion in the 
small parameter $1/g$ \cite{vanRos95}. The results are compared to 
calculations for a plane wave since $d\, \ll\, L$ and there is a nearly 
complete mixing of modes in the sample, giving a uniform average intensity 
along a cross section of the sample. The agreement between theory and 
experiment is surprising, however, since $1/g \gtrsim 0.1$ for all samples, 
reaching a value of approximately 0.3 for sample c, and is by no means 
small. Furthermore, the influence of absorption was not included in the 
calculations, whereas the samples used in the experiment are strongly 
absorbing with $L > L_a \approx 40$ cm, where $L_a$ is the exponential 
absorption length \cite{Gen93a}.

We now consider the full transmission distribution. The theoretical 
expressions for the full distribution function in Refs.\ 
\cite{vanRos95,Kog95,vanLang96} are given as functions of $g$ for 
nonabsorbing samples. In the present case of strong absorption, the 
photon number is not conserved, and $g$ cannot be defined in terms of 
the steady state transmission, while serving as a useful measure of the 
proximity to the localization transition. This can be seen by noting that 
the reduction of the average transmission due to absorption would lead to 
a reduced value of $g$ even though the presence of absorption tends to 
lessen the degree of correlation in the sample and to push the system 
farther from the localization threshold. On the other hand, a parameter 
which characterizes the transmission distribution as well as the closeness 
to the localization threshold, even in the presence of absorption, is the 
degree of correlation of intensity in different coherence areas of the 
transmitted speckle pattern, $<\!\!\delta s_{ab}\delta s_{ab'}\!\!>$.  Were 
this correlation to vanish, fluctuations in different coherence areas would 
be independent and the transmission distribution would be Gaussian by the 
central limit theorem with $var(s_a)\,  \equiv\,  <\!\!s_a^2\!\!>_c = 1/N$. 
As a result of nonlocal correlation, however, the variance of the 
transmission is enhanced. It is given by $<\!\!s_a^2\!\!>_c\, =\, 
(<\!\!s_{ab}^2\!\!>_c - 1)/2\, =\, <\!\!\delta s_{ab}\delta s_{ab'}\!\!>$ \cite{Kog95,Feng88,Mello88,Gar93}. The last equality is consistent with the 
results of Ref.\ \cite{Gar93} when the cumulant intensity correlation 
function is properly normalized to the renormalized average transmission 
\cite{Shap97}. In that case, the crossing parameter found by Shnerb and 
Kaveh \cite{Shn91} which determines the intensity distribution is found 
experimentally to be equal to $<\!\!\delta s_{ab}\delta s_{ab'}\!\!>$ 
\cite{Gen93,Gar93}. The connection of $<\!\!\delta s_{ab}\delta s_{ab'}\!\!>$ 
to the full transmission distribution can be seen by considering the 
expression of Refs.\ \cite{vanRos95,Kog95} for $P(s_a)$ in the absence 
of absorption in the limit $g\ \gg\ 1$,
\begin{equation}
\label{psa}
P(s_a) = \int_{-i\infty}^{+i\infty}\frac{dx}{2\pi i}exp[xs_a - \Phi(x)],
\end{equation}
where \[\Phi(x) = g ln^2(\sqrt{1 + x/g} + \sqrt{x/g})\] is the generating 
function. From Eq.\ (\ref{psa}) one obtains the expression for 
$<\!\!s_a^2\!\!>_c$ in terms of $g$, 
\begin{equation}
\label{secondc}
<\!\!s_a^2\!\!>_c = \frac{2}{3g}.
\end{equation}
From these expressions, a general relation for $P(s_a)$ in terms of 
$<\!\!s_a^2\!\!>_c$, or equivalently $<\!\!\delta s_{ab}\delta s_{ab'}\!\!>$, 
can be found by using Eq.\ (\ref{secondc}) to define a new parameter $g' = 2/3\!\!<\!\!s_a^2\!\!>_c$ which is substituted for $g$ into Eq.\ (\ref{psa}). 
Plots of $P(s_a)$ obtained by following this procedure with $g'$ determined 
from the measured values of $<\!\!s_a^2\!\!>_c$ are shown as the solid lines 
in Figs. 1 and 2. We find that $P(s_a)$ is accurately given even for the 
lowest value of $g'$ of 3.06 (sample c).  The distribution of Eq.\ (\ref{psa}) 
with $g'$ substituted for $g$ gives the exponential tail, $P(s_a) \sim 
exp(-g's_a)$ in the limit $s_a \gg 1$. For $s_a\, \geq\, 2.0$, the linear 
fit to the logarithm of the measured transmission distribution for sample c 
gives a slope of $2.71\, \pm\, 0.06$ in accord with the exponential fit of 
the theoretical curve of 2.70 in this range and is quite close to its 
predicted asymptotic value of 3.06 for $s_a \gg 1$.

The extent of the agreement of Eq.\ (\ref{psa}) when $g'$ is substituted for 
$g$ can also be gauged from the comparison between the calculated (circles) 
and the measured (squares) moments of the transmission distribution shown in 
Fig. 4 for samples with $g' = 10.2\, \pm\, 0.1$ and $g' = 3.06\, \pm\, 0.04$. 
The moments calculated from the theory are close to those obtained from the 
measured distributions. At $n\, =\, 10$, these defer by approximately 
$10\, \%$ which is within the experimental error. Thus it appears that 
$P(s_a)$ can be expressed as a function of the parameter $<\!\!s_a^2\!\!>_c$.

The agreement between theory and experiment indicates that the ratio of 
moments is accurately reflected in Eq.\ (\ref{psa}). The dependence of the 
second cumulant itself upon sample dimensions is shown in Fig. 5. In the 
limit $g \gg 1$, in the absence of absorption, \mbox{$<\!\!s_a^2\!\!>_c\, =\, 
2L/3N\ell$}. The straight line in the figure is drawn through the first data 
point and represents $<\!\!s_a^2\!\!>_c\, \sim\, L/N$. As $g \rightarrow 1$, 
and the localization threshold is approached, the scaling theory of 
localization \cite{gang4} suggests that $g$ falls more rapidly and hence 
$<\!\!s_a^2\!\!>_c$ should increase superlinearly with sample length.  
Instead, we find that $<\!\!s_a^2\!\!>_c$ depends sublinearly upon $L$. 
This is presumably due to the presence of absorption which diminishes the 
degree of nonlocal correlation. This raises the question of whether the 
transmission distribution continues to broaden as $L$ increases or, instead, 
it reaches a limiting distribution for particular sample parameters. 

In conclusion, we have measured the total transmission distribution of 
microwave radiation in quasi-one-dimensional absorbing samples with small 
values of $g$. We find that the distribution can be described using an 
expression originally derived for nonabsorbing samples in the limit 
$g \gg 1$ when this expression is reformulated as a function of the single 
parameter $g'\, =\, 2/3<\!\!s_a^2\!\!>_c$ determined from the measurements. 
The validity of the expression for values of $g'$ as small as 3, well beyond 
the limits assumed in the calculations, may well be associated with the 
identification of $<\!\!s_a^2\!\!>_c$ with 
$<\!\!\delta s_{ab}\delta s_{ab'}\!\!>$, the degree of spatial correlation 
in the sample, which is the key microscopic parameter in mesoscopic physics.

We are pleased to acknowledge stimulating discussions with E. Kogan, M. C. 
W. van Rossum and B. Shapiro. We thank E. Kuhner, Z. Ozimkowski and D. 
Genack for constructing and testing the integrating sphere as well as W. 
Polkosnik for his help in automating the experiment. We are grateful to N. 
Garcia for his encouragement and fruitful discussions. This work was 
supported by a National Science Foundation Grant No. 9632789 and by a PSC-
CUNY award.

\pagebreak
\noindent
{\bf FIGURES}:\\

\noindent
Fig. 1. Distribution function of the normalized transmission $P(s_a)$ for 
three samples with dimensions (a) $d = 7.5$ cm, $L = 66.7$ cm, (b) $d = 5.0$ 
cm, $L = 50.0$ cm, and (c) $d = 5.0$ cm, $L = 200$ cm.\\

\noindent
Fig. 2. Semilogarithmic plot of the  transmission distributions for the
same samples as in \mbox{Fig.\ 1}.\\

\noindent
Fig. 3. Plot of $<\!\!s_a^3\!\!>_c$ versus $<\!\!s_a^2\!\!>_c^2$. The solid 
line represents a least square linear fit to the data.\\

\noindent
Fig. 4. Comparison of the calculated (circles) and measured (squares) moments 
of the transmission distribution for samples with (a) $g' = 10.2$ and (b) 
$g' = 3.06$.\\

\noindent
Fig. 5. Dependence of the second cumulant $<\!\!s_a^2\!\!>_c$ upon sample 
dimensions.\\


\begin{thebibliography}{99}
\bibitem{Sheng90} {\it The Scattering and Localization of Classical Waves}, 
edited by P. Sheng (World Scientific, Singapore, 1990).
\bibitem{Alt91} {\it Mesoscopic Phenomena in Solids}, edited by B. L. 
Altshuler, P. A. Lee, and R. A. Webb (North-Holland, Amsterdam, 1991).
\bibitem{vanRos95} Th. M. Nieuwenhuizen and M. C. W. van Rossum, Phys. Rev. 
Lett. {\bf 74}, 2674 (1995).
\bibitem{Kog95} E. Kogan and M. Kaveh, Phys. Rev. B {\bf 52}, R3813 (1995).
\bibitem{Land70} R. Landauer, Phil. Mag. {\bf 21}, 863 (1970).
\bibitem{Steph87} M. J. Stephen and G. Cwilich, Phys. Rev. Lett. {\bf 59}, 
285 (1987).
\bibitem{Lee85} P. A. Lee and A. D. Stone, Phys. Rev. Lett. {\bf 55}, 1622 
(1985).
\bibitem{Alt85} B. L. Altshuler and D. E. Khmelnitskii, Sov. Phys. JETP Lett. 
{\bf 61}, 359 (1985).
\bibitem{Feng88} S. Feng, P. A. Lee, and A. D. Stone, Phys. Rev. Lett. 
{\bf 61}, 834 (1988).
\bibitem{Mello88} P. A. Mello, E. Akkermans, and B. Shapiro, Phys. Rev. 
Lett. {\bf 61}, 459 (1988).
\bibitem{Gar89} N. Garcia and A. Z. Genack, Phys. Rev. Lett. {\bf 63}, 
1678 (1989).
\bibitem{Gen93} A. Z. Genack and N. Garcia, Europhys. Lett. {\bf 21}, 
753 (1993).
\bibitem{deBoer94} J. F. de Boer, M. C. W. van Rossum, M. P. van Albada, 
Th. M. Nieuwenhuizen, and A. Lagendijk, Phys. Rev. Lett. {\bf 73}, 2567 
(1994).
\bibitem{vanLang96} S. A. van Langen, P. W. Brouwer, and C. W. J. Beenakker, 
Phys. Rev. E {\bf 53}, 1344 (1996).
\bibitem{Gen93a} A. Z. Genack, N. Garcia, and A. A. Lisyansky, in {\it 
Photonic Band Gaps and Localization}, edited by C. M. Soukoulis (Plenum 
Press, New York, 1993). 
\bibitem{Gar93} N. Garcia, A. Z. Genack, R. Pnini, and B. Shapiro, Phys. 
Lett. A {\bf 176}, 458 (1993).
\bibitem{Shap97} B. Shapiro, privite communication.
\bibitem{Shn91} N. Shnerb and M. Kaveh, Phys. Rev. B {\bf 43}, 1279 (1991).
\bibitem{gang4} E. Abrahams, P. W. Anderson, D. C. Licciardello, and T. V. 
Ramakrishnan, Phys. Rev. Lett. {\bf 42}, 673 (1979). 
\end{thebibliography}
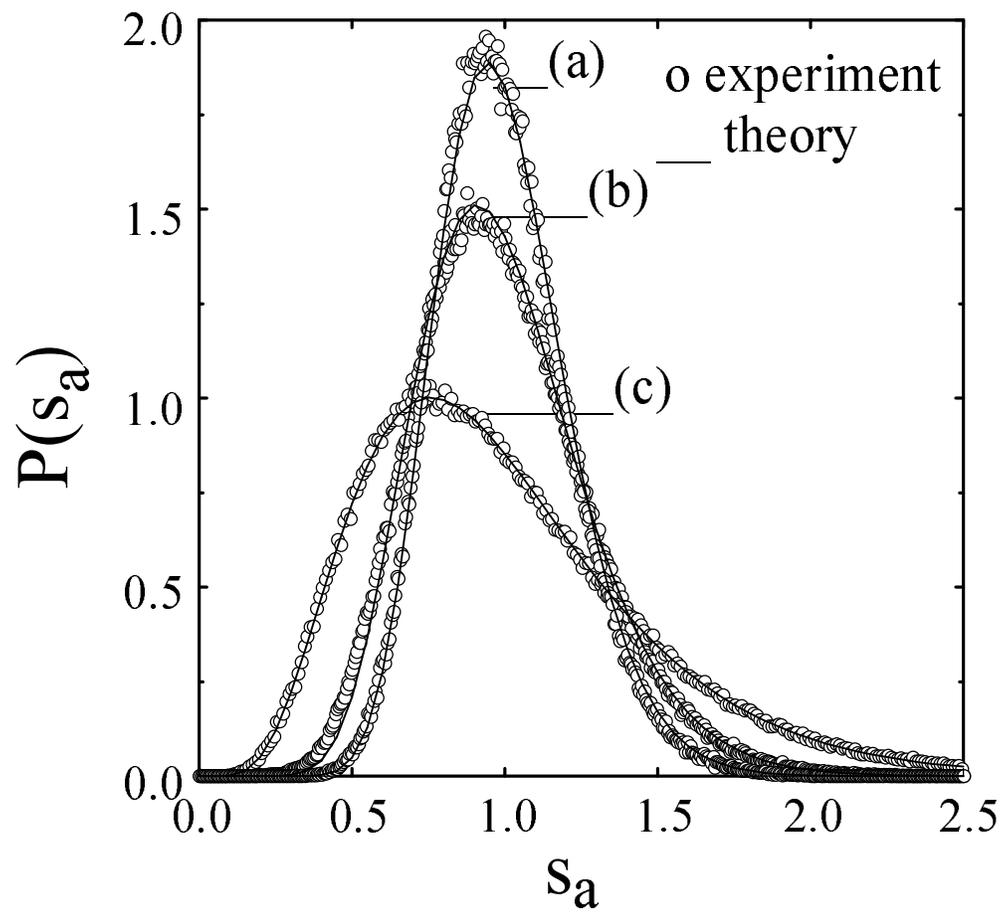
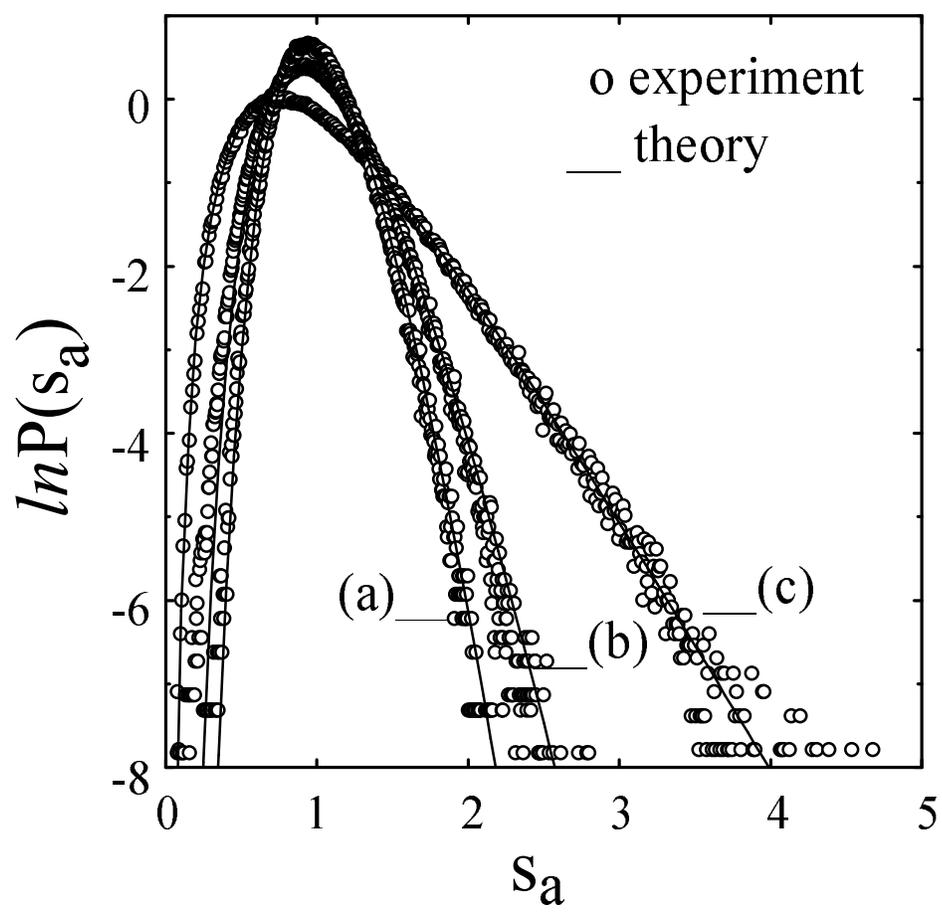
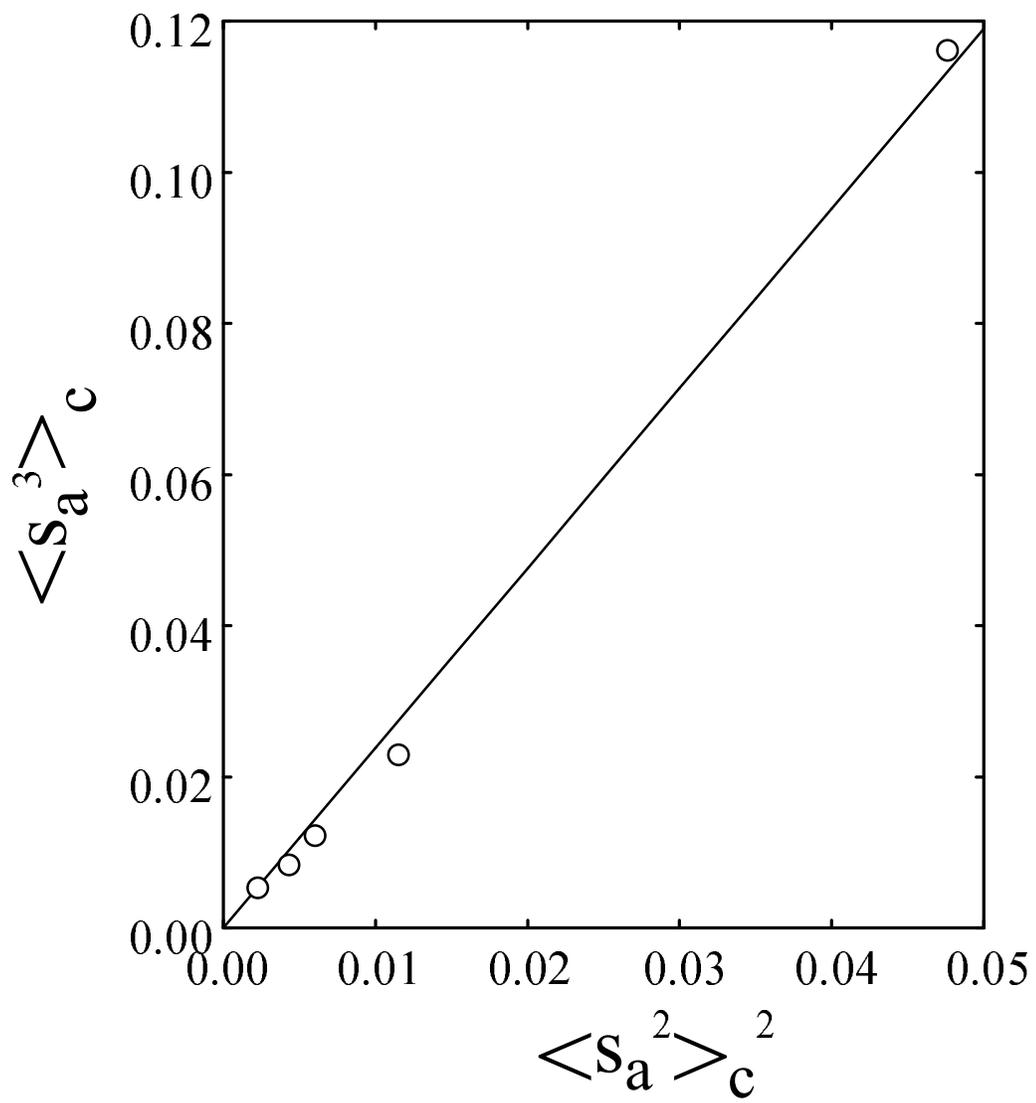
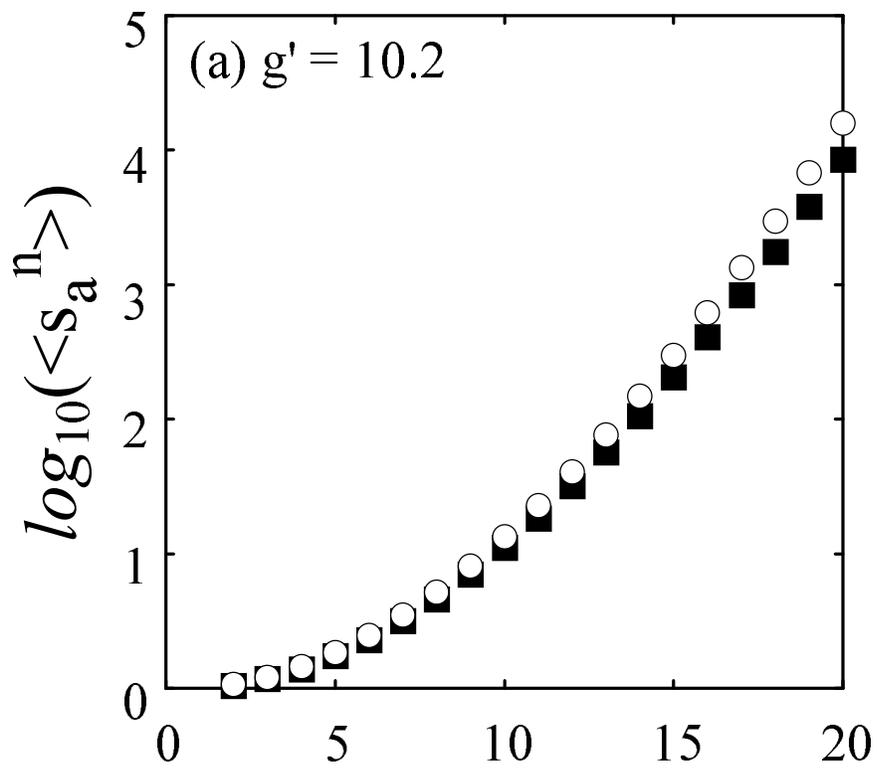
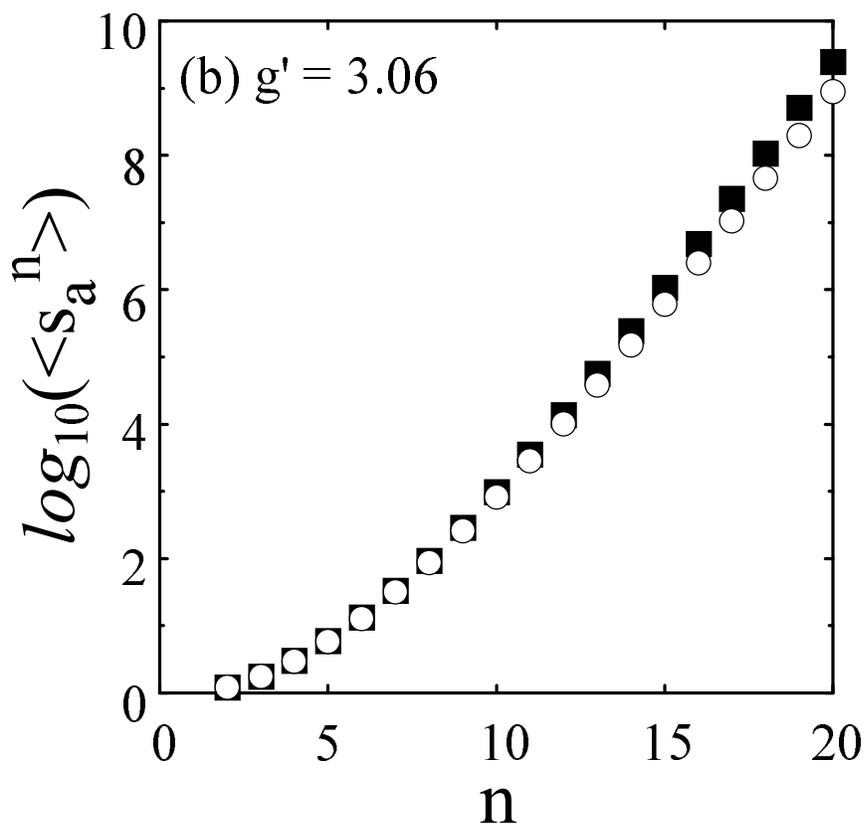
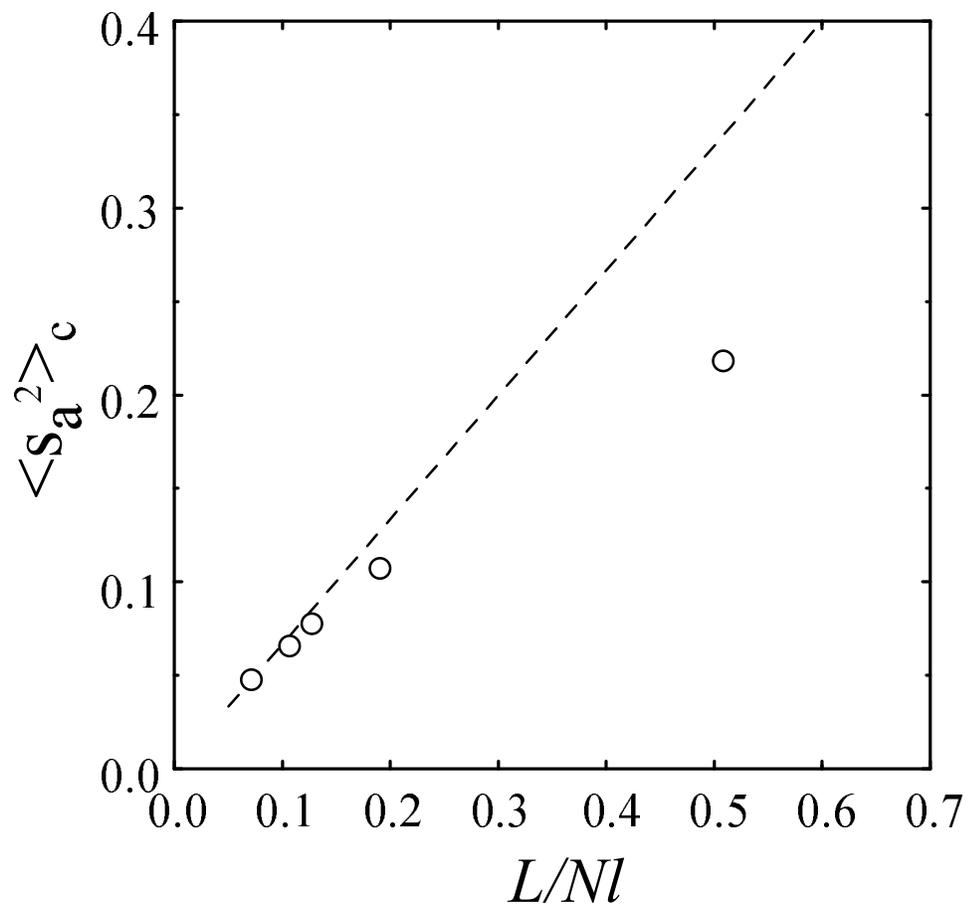
\end{document}